\documentclass[traditabstract]{aa}
\usepackage[utf8]{inputenc}
\usepackage[colorlinks]{hyperref}
\usepackage{amsmath}
\usepackage{amssymb}
\usepackage{microtype}
\usepackage[T1]{fontenc}
\usepackage{lmodern}
\usepackage{natbib}
\usepackage{graphicx}
\usepackage{microtype}
\usepackage{color}
\usepackage{multirow}
\usepackage{xspace}
\bibpunct{(}{)}{;}{a}{}{,} % to follow the A&A style
\newcommand{\src}{Swift~J1845.7$-$0037\xspace}
\newcommand{\NS}{\emph{NuSTAR}\xspace}

\begin{document}

\title{First characterization of \src with \NS}
\author{V.\,Doroshenko\inst{1,3}, S.\,Tsygankov\inst{2,3}, J.\,Long\inst{1},  A.\,Santangelo\inst{1}, S.~Molkov\inst{3}, A.~Lutovinov\inst{3,4}, L.D.~Kong\inst{5,6}, S.\,Zhang\inst{5}}  
\institute{Institut für Astronomie und Astrophysik, Sand 1, 72076 Tübingen, Germany}

\institute{Institut für Astronomie und Astrophysik, Sand 1, 72076 Tübingen, Germany\and
Department of Physics and Astronomy, FI-20014 University of Turku, Turku, Finland \and
Space Research Institute of the Russian Academy of Sciences, Profsoyuznaya Str. 84/32, Moscow 117997, Russia \and
Moscow Institute of Physics and Technology, Moscow region, Dolgoprudnyi, Russia \and 
Key Laboratory for Particle Astrophysics, Institute of High Energy Physics, Chinese Academy of Sciences, 19B Yuquan Road, Beijing 100049, People’s Republic of China\and
University of Chinese Academy of Sciences, Chinese Academy of Sciences, Beijing 100049, People’s Republic of China
}

\abstract{
The hard X-ray transient source \src was discovered in 2012 by \textit{Swift}/BAT. However, at that time no dedicated observations of the source were performed. On Oct 2019 the source became active again, and X-ray pulsations with a period of $\sim199$\,s were detected with \textit{Swift}/XRT. This triggered follow-up observations with \NS. Here we report on the timing and spectral analysis of the source properties using \NS and \textit{Swift}/XRT. The main goal was to confirm pulsations and search for possible cyclotron lines in the broadband spectrum of the source to probe its magnetic field. Despite highly significant pulsations with period of 207.379(2) were detected, no evidence for a cyclotron line was found in the spectrum of the source. We therefore discuss the strength of the magnetic field based on the source flux and the detection of the transition to the ``cold-disc'' accretion regime during the 2012 outburst. Our conclusion is that, most likely, the source is a highly magnetized neutron star with $B\gtrsim10^{13}$\,G at a large distance of $d\sim10$\,kpc. The latter one  consistent with the non-detection of a cyclotron line in the \NS energy band.
}
\authorrunning{V. Doroshenko et al.}
\maketitle

\section{Introduction}
The transient source \src was discovered in the XMM slew survey as XMMSL1~J184555.4-003941 \citep{2008A&A...480..611S}, and was later identified as a hard X-ray transient following the detection by \emph{Swift}/BAT in 2012 \citep{2012ATel.4130....1K}. In 2019 the source was detected by MAXI \citep{2019ATel13189....1N} and followed-up by \textit{Swift}/XRT, which allowed the detection of pulsations at $\sim200$\,s \citep{2019ATel13195....1K} and better constrained the X-ray position of the source thereby suggesting 2MASS\,J18455462-0039341 as optical counterpart candidate  \citep{atel13218,atel13219}. An analysis of the spectral energy distribution of this object suggests it is a strongly absorbed early type star with $T_{\rm eff}\sim28000-33000$\,K \citep{atel13211,atel13222} consistent with the expected Be-transient origin of the source \citep{2019ATel13195....1K}.
Here we report results of the follow-up observations of \src with the \NS observatory \citep{2013ApJ...770..103H} aimed at provide first  characterization of its broadband X-ray properties. 

\section{Observations and data analysis}

Following the detection of pulsations with MAXI \citep{2019ATel13195....1K} we triggered an observation of \src with \textit{NuSTAR} on MJD~58777 for 44\,ks (Obsid. 90501347002, effective exposure $\sim23.5$\,ks). Data reduction was carried out using the HEASOFT~6.26.1 package 
with current calibration files (CALDB version 20191008) and standard data reduction procedures as described in the instruments documentation. Source spectra were extracted from a region of 54$^{\prime\prime}$ radius around the position of \src, whereas background spectra were extracted from a circular region of 147$^{\prime\prime}$ radius away from the source. The extraction regions were optimized to increase the signal to noise ratio in the hard energy band $\ge40$\,keV as described in \citep{2018A&A...610A..88V}. The spectra for the two \NS modules (FPMA and FPMB) were extracted independently and modeled simultaneously in the 3$-$79\,keV range. All spectra were grouped to include at least 25 counts per energy bin. Lightcurves of the two modules were co-added and background subtracted unless stated otherwise. We also analyzed \textit{Swift}/XRT observations for which the source spectrum was extracted using online 
tools\footnote{\url{http://www.swift.ac.uk/user_objects/}} provided by the UK Swift Science Data Centre \citep{2009MNRAS.397.1177E}. The spectrum was grouped to contain at least 25 counts per energy bin and modeled simultaneously with the \NS spectra using the {\sc xspec} 12.10.1.f package \citep{1996ASPC..101...17A}.

\subsection{Timing analysis}

\begin{figure}
    \includegraphics[width=\columnwidth]{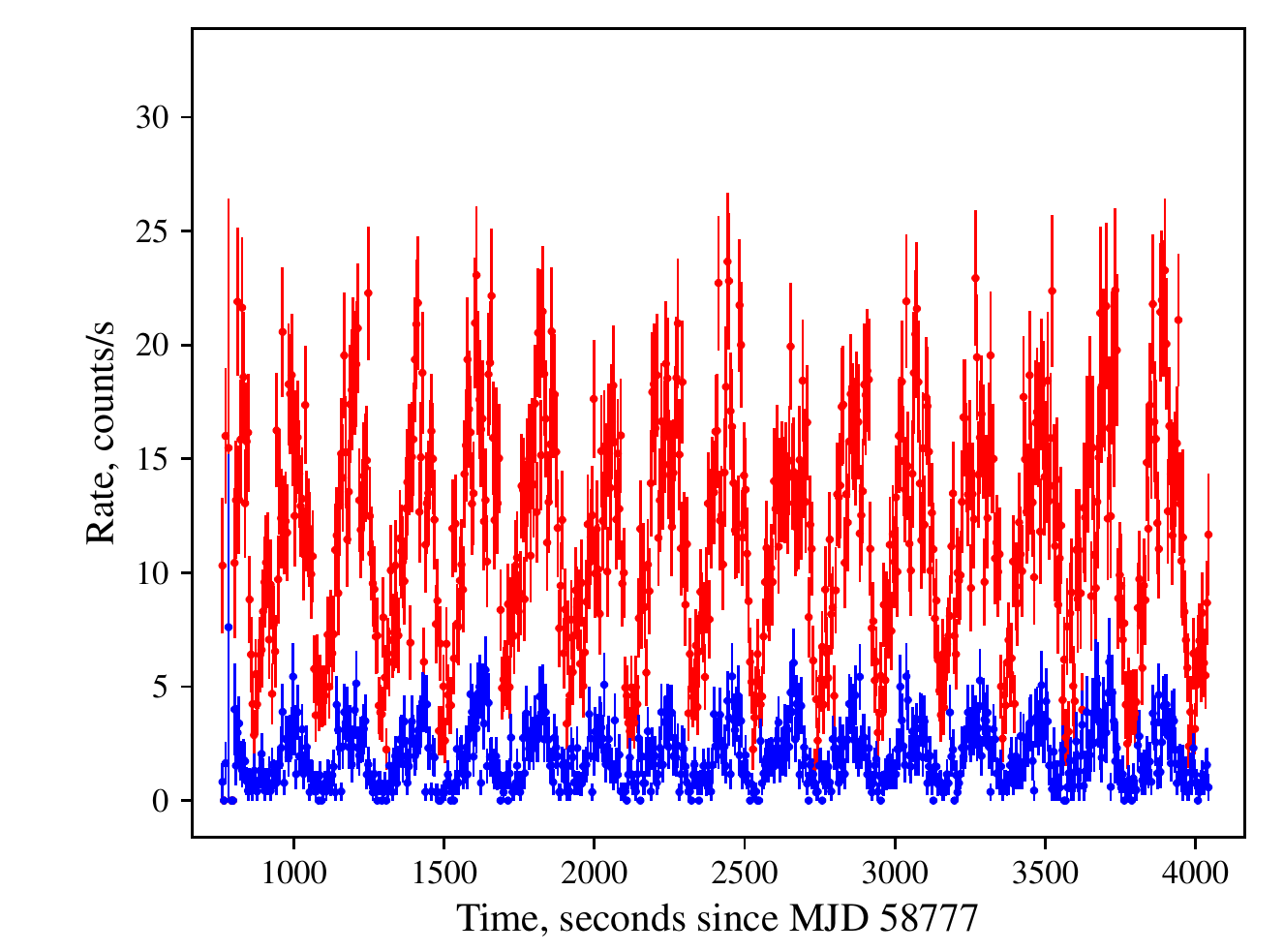}
        \caption{Part of the lightcurve of the source in the 3-10\,keV (red) and 20-40\,keV (blue) bands as observed by \NS. Strong pulsations with a period of $\sim200$\,s are observed throughout the whole energy range.}
    \label{fig:nulc}
\end{figure}

As shown in Fig.~\ref{fig:nulc}, strong pulsations with a period of $\sim200$\,s are clearly visible directly in the source lightcurve. Epoch folding search reveals a strong peak around $207.4$\,s. We assumed the source position reported by \cite{2019ATel13195....1K} for the barycentric correction. No binary correction was applied as the orbital parameters are not known. To refine this value and estimate the uncertainty of the measured spin period, we used a phase-connection technique \citep{1981ApJ...247.1003D}. Times of arrival (TOAs) of individual pulse cycles were determined by direct fitting the average template obtained by folding the entire 3-20\,keV lightcurve with the period found above, allowing the shifting and rescaling of the template in time and count-rate by arbitrary factors considered as free parameters. The average uncertainty of TOAs was found to be $\sim1.7$\,s. The individual TOAs were then fitted assuming a constant period, i.e., $t_{i,calc}=n_i\times p+t_0$, which resulted in the period estimate of $P=207.379(2)$. No evidence for the change of the spin period within the observation was found, which is not surprising considering its relatively short duration.

\begin{figure}
    \includegraphics[width=\columnwidth]{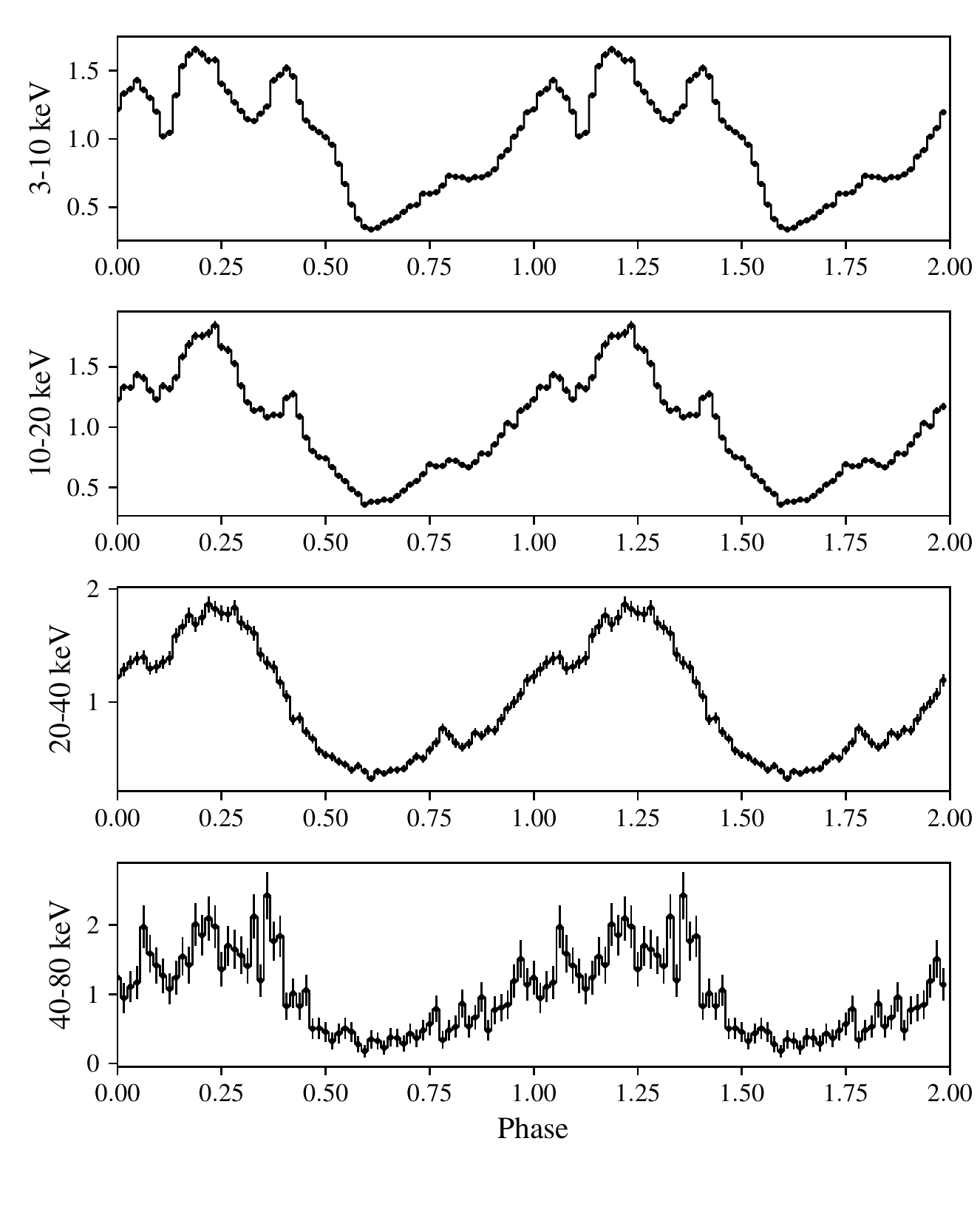}
    \includegraphics[width=\columnwidth]{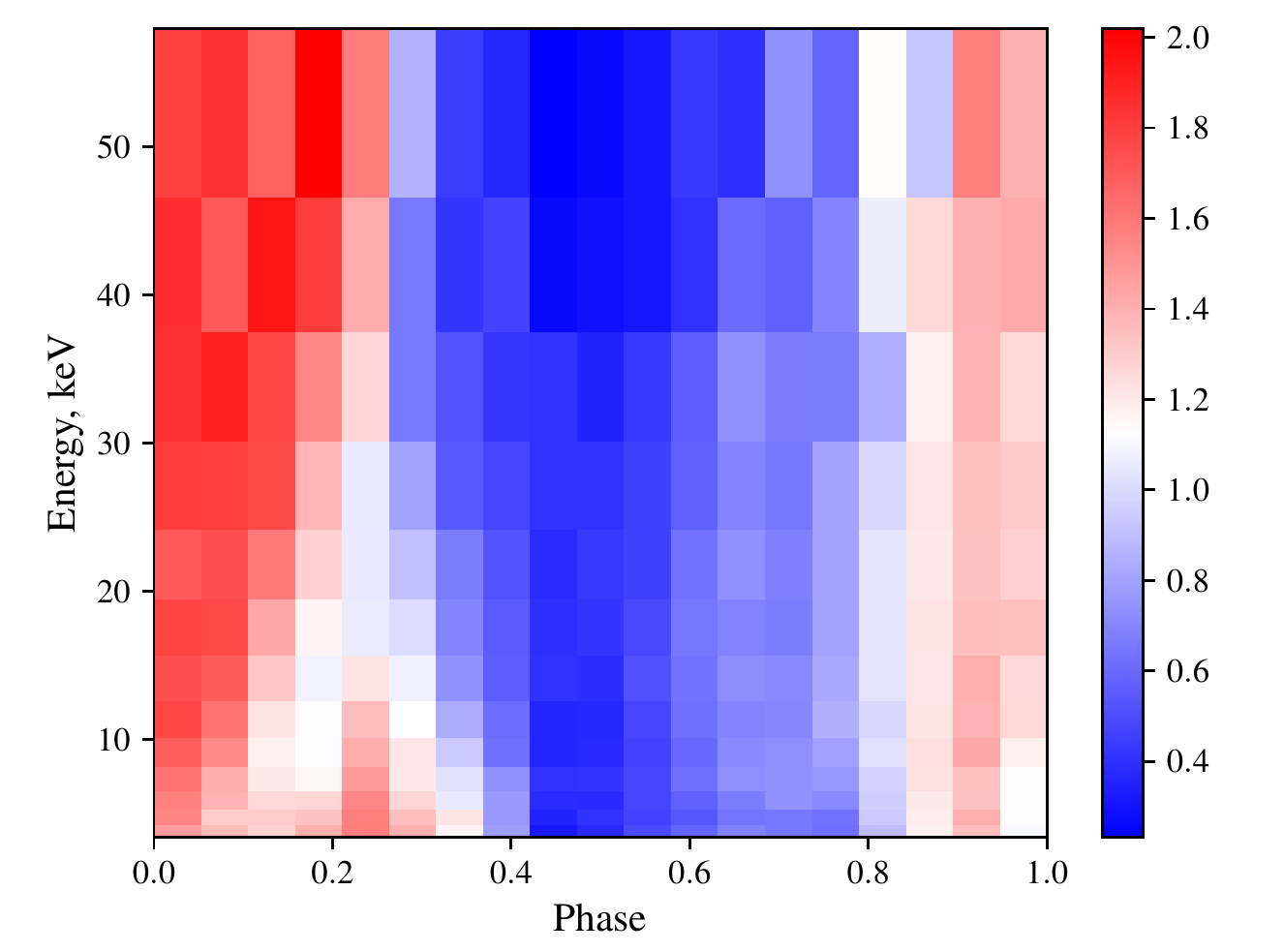}
        \caption{Representative pulse profiles of the source as observed by \NS in several energy bands (top), and phase-energy matrix showing the energy evolution of the (normalized) pulse profiles in details (bottom). In the later case, slices along the constant energy represent normalized pulse profiles similar to those shown in the top panel. Note the energy dependence of pulse profile shape, particularly around $\sim15$\,keV.}
    \label{fig:pps}
\end{figure}

\begin{figure}
    \includegraphics[width=\columnwidth]{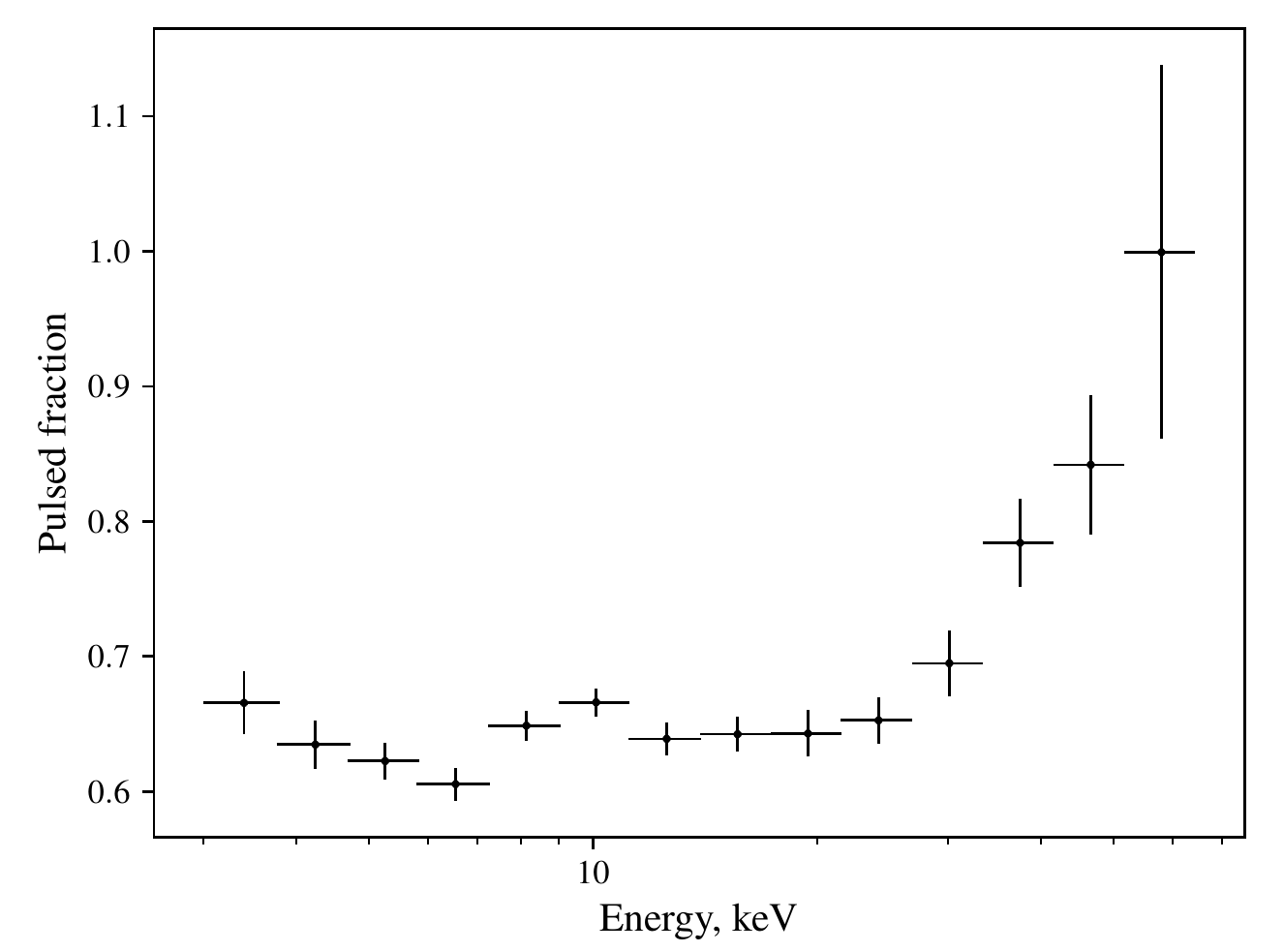}
        \caption{Pulsed fraction of \src defined as $(max(rate)-min(rate))/((max(rate)+min(rate))$ as a function of the energy.}
    \label{fig:pf}
\end{figure}

Using the obtained period, we folded the lightcurves extracted in several energy bands in order to investigate the evolution of the pulse profile with the energy. As shown in Fig.~\ref{fig:pps}, the pulse profile at lower energies exhibits a complex morphology, showing several narrower peaks within a broader main peak and its rising phase. At higher energies ($\gtrsim15$\,keV) these structures gradually disappear. This can be better illustrated with the phase-energy matrix shown in the same figure. In addition, the pulsed fraction 
decreases with the energy at soft X-rays, whereas increases at harder energises reaching almost 100\% at the highest energies (Fig.~\ref{fig:pf}), as typical for bright X-ray pulsars \citep{2009AstL...35..433L}. 
The observed pulse profile evolution suggests a possible presence of an independent soft spectral component which exhibits a different and more complex dependence on the pulse phase.

\subsection{Spectral analysis}

As reported in \citep{atel13208}, the source spectrum measured with \NS is typical of X-ray pulsars and can be described with a Comptonization model with an electron temperature of $6-7$\,keV. It does not show any evident absorption features suggestive of electron cyclotron resonance scattering lines. This conclusion does not depend on the model chosen to describe the continuum emission. The latter can be described by different continuum models such as, for example, a cutoff power law combined with partial-covering absorption or an additional soft blackbody component, a combination of different Comptonization models, etc.).  
Considering that the interpretation of parameters of commonly used phenomenological models is often rather ambiguous, we adopt here one of the simplest commonly used model adequately describing the broadband phase-averaged spectrum of the source. This is an absorbed power law with the Fermi-Dirac \citep{1986LNP...255..198T} cutoff implemented using the \texttt{mdefine} command in the {\sc xspec} package. The absorption was modeled using \texttt{TBabs} model and abundances from \citet{2000ApJ...542..914W}.

\begin{figure}
    \includegraphics[width=\columnwidth]{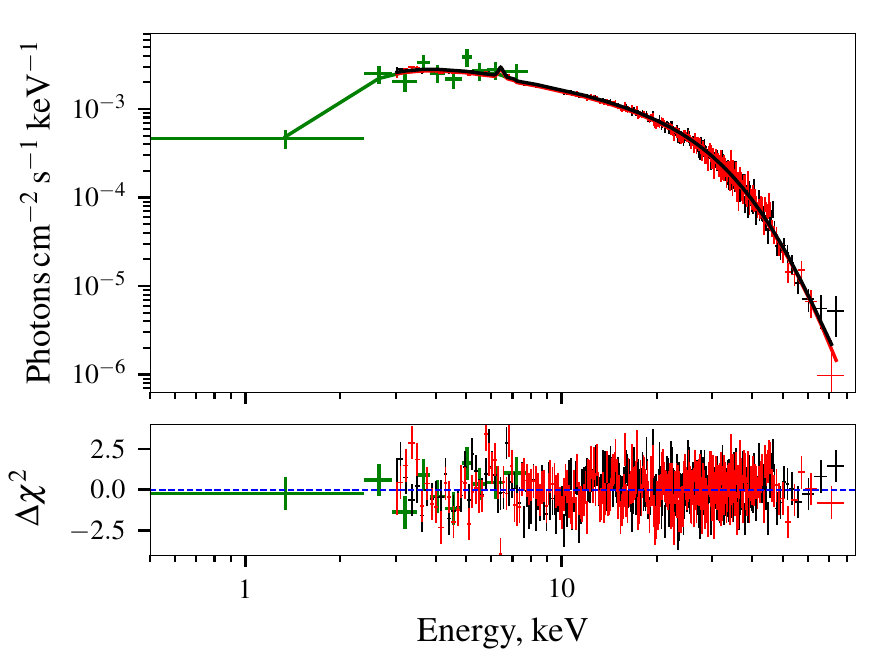}
        \caption{Unfolded spectrum assuming the best-fit \texttt{FDCUT} continuum as observed by \textit{Swift}/XRT (green), and \NS (black for FPMA and red for FPMB). The lower panel shows corresponding residuals.}
    \label{fig:spe}
\end{figure}

Besides the continuum we also included in the model a narrow Gaussian line at $\sim6.4$\,keV to account for the weak iron line observed in the spectrum of the source, and cross-normalization constants to account for slight differences in the absolute flux between two \textit{NuSTAR} telescope modules and \textit{Swift}/XRT. The energy resolution of \NS does not allow to constrain the iron line width, so it was fixed to 0.1\,keV. For the phase-averaged analysis we used only a single XRT observation (Obsid. 00032472018), the closest in time to the \NS observation. In fact, \textit{Swift} observed the source on MJD~58775.5 whereas the \NS observation started 1.5 days later. This resulted in a much lower exposure ($\sim750$\,s) and counting statistics in the soft energy band. Still, extending the lower energy coverage helped to constrain the absorption column which appears to be a factor of two higher compared to the expected interstellar value ($\sim1.72\times10^{22}$\,atoms\,cm$^{-2}$ \citealt{2016A&A...594A.116H}).  As we said, no evidence for a cyclotron line was obtained (Fig.~\ref{fig:spe}). The best-fit parameters for the continuum component and the iron line are listed in Table~\ref{tab:pars}. In addition the table includes the values of the cross-normalization constants, and the observed and unabsorbed fluxes in the 0.1-100\,keV energy band.

\begin{figure}
    \includegraphics[width=\columnwidth]{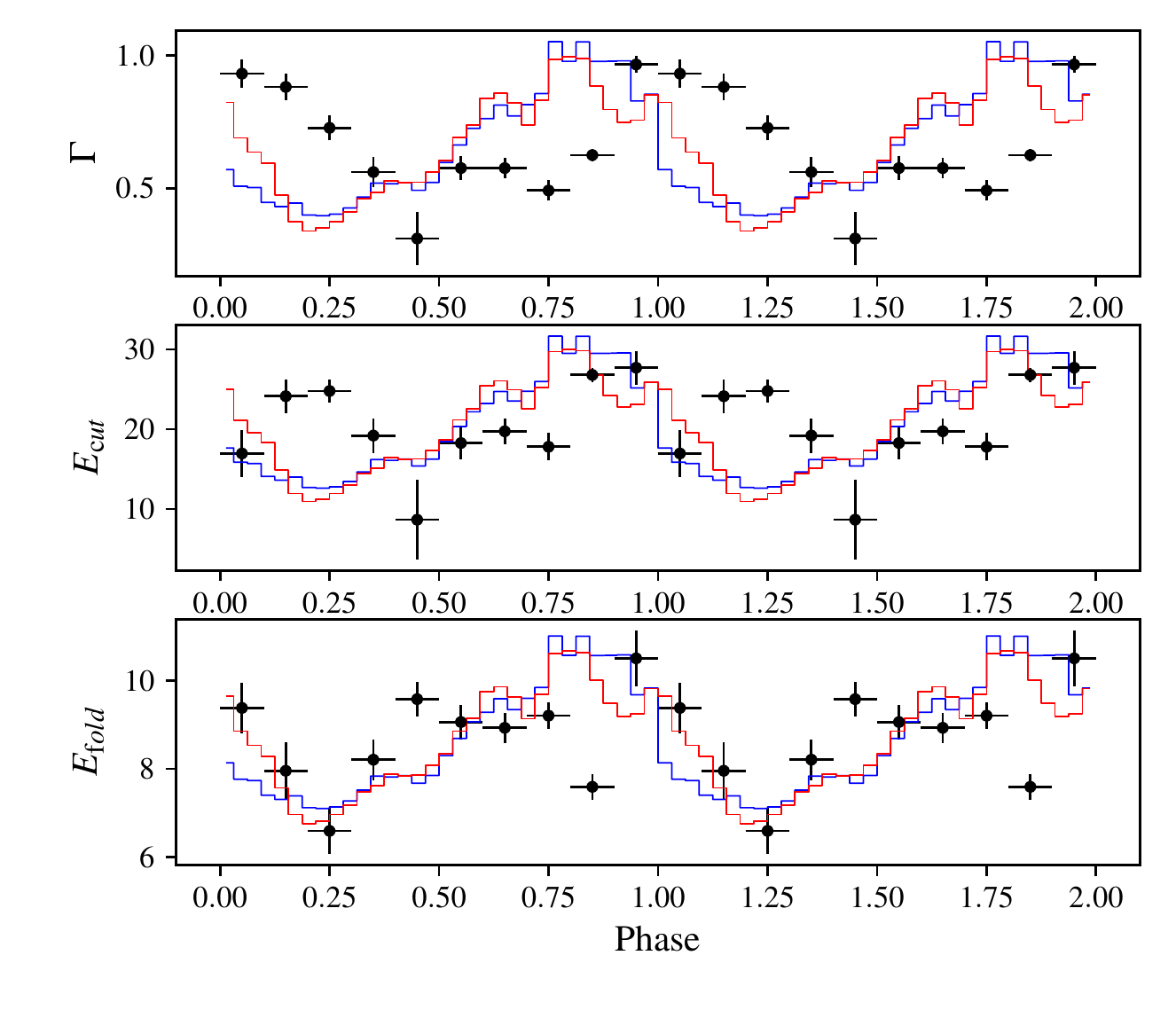}
        \caption{Evolution of the \texttt{FDCUT} continuum parameters with the pulse phase. The pulse profiles of the source in soft (3-20\,keV, red) and hard (20-80\, keV, blue) are also plotted for reference.}
    \label{fig:phres}
\end{figure}

We have also performed pulse phase resolved spectral analysis using the model described above. \textit{Swift}/XRT data could not be used. We fixed, therefore, the absorption column and iron line parameters to values derived from the phase-averaged spectral analysis. Variations of the parameters with the pulse phase are shown in Fig.~\ref{fig:phres}.  

\begin{table}[t]
    \centering
    \caption{Best-fit parameters of the phase-averaged spectrum of \src using the \texttt{FDCUT} model. Cross-normalization constants relative to \texttt{FPMA}, observed and intrinsic fluxes in 0.1-100\,keV band, equivalent width of the iron line $W_{Fe}$, and fit statistics are also reported.}
    \begin{tabular}{ll}
         Parameter & Value \\
         \hline\hline
         $N_H$, $10^{22}$ atoms\,cm$^{-2}$ & 5.4(4) \\
         $\Gamma$ & 0.69(2) \\
         $A_\Gamma$, ph\,keV$^{-1}$cm$^{-2}$\,s$^{-1}$ & 0.0102(4) \\
         $E_{\rm cut}$, keV & 22.4(6) \\
         $E_{\rm fold}$, keV & 8.7(2) \\
         $E_{Fe}$, keV & 6.44(7) \\
         %$A_{Fe}$ ph\,cm$^{-2}$\,s$^{-1}$ & 0.00017(2) \\
         $W_{Fe}$, keV & 0.0766(6) \\
         $C_{\rm FPBB} $ & 1.019(4) \\
         $C_{\rm XRT}$ & 1.13(7) \\
         $F_{\rm x, obs}$, 10$^{-10}$\,erg\,cm$^{-2}$\,s$^{-1}$ & 7.61\\
         $F_{\rm x, unabs}$, 10$^{-10}$\,erg\,cm$^{-2}$\,s$^{-1}$ & 8.23\\
         $\chi^2$/dof & 607.16/585 \\
         \hline
    \end{tabular}
    
    \label{tab:pars}
\end{table}

\section{Discussion and summary}
The transient source \src was discovered in 2012. No dedicated observations in the broad energy band were performed and the source was only observed by all sky monitors and \textit{Swift}/XRT. 
In Oct~2019 a new outburst from the source was observed by MAXI \citep{2019ATel13191....1K}, which was followed-up first with \textit{Swift}, which provided hints for pulsations \citep{2019ATel13195....1K} and then with \NS. We observed the source on Oct 21, 2019, confirming strong pulsations with high significance. 
Besides confirming the nature of the source, one of the prime goals of our investigation was to search for possible electron resonance scattering cyclotron lines in the spectrum, to estimate the strength of the pulsar's magnetic field. Unfortunately, our analysis did not show evidence for cyclotron lines. In fact, the broadband continuum spectrum of the source can be well described by different phenomenological models without the necessity to include any absorption line in any of the models. The source's spin period remained constant, which, together with the unknown orbit and distance, precludes any estimate of the magnetic field by modeling spin evolution with accretion torque models \citep[see, e.g.,][for the current results]{2016A&A...593A..16T,2017A&A...605A..39T,2018A&A...613A..19D,2019MNRAS.485..770L}.

Still, important information on the source properties has been deduced based the X-ray phenomenology. First of all, the comparison of the source's flux with typical luminosities ( $\sim10^{37}$\,erg\,s$^{-1}$) reached in outbursts by other \textit{Be} transients, suggests a large distance to the source, i.e., $\sim10$\,kpc. This conclusion is consistent with the observed strong absorption both in the X-ray (see Table 1) and optical bands \citep{atel13222}, which actually exceeds the expected interstellar absorption integrated over the entire Galaxy, and strong local absorption is not common for BeXRBs. 
We note that this conclusion is in line with the observed complex energy evolution of the pulse profile and observed variation of spectral parameters with pulse phase.

Indeed, strong energy dependence of the pulse profiles suggests a strong energy dependence of the intrinsic beam pattern as expected for high luminosity accreting pulsars, where an accretion column is expected to develop \citep{1976MNRAS.175..395B}. On the contrary, pulse profiles of complex morphology and energy dependence are normally not observed in accreting pulsars of lower luminosity ($\sim5\times10^{36}$\,erg\,s$^{-1}$, \citealt{2007PASJ...59..961K}). The complex phase dependence of the source spectrum can be explained by the phase-dependent visibility of different regions of the column and of parts of the neutron star's surface illuminated by the column \citep{2003ApJ...590..424K,2013ApJ...777..115P}.
We emphasize that any meaningful interpretation of the observed variations of spectra parameters of the phenomenological models like \texttt{FDCUT} is not straightforward, especially considering that the observed pulse profile evolution with the energy suggest a combination of emission components from the two poles of the pulsar, that may contribute to the spectrum at any phase, although with different weight. Still, one might notice that variation of the photon index $\Gamma$ and folding energy $E_{\rm fold}$ follows different patterns. Whereas the latter traces the overall flux evolution the former is phase shifted by $\sim0.25$. This could again point to the presence of two independent spectral components with a similar visibility pattern throughout the pulse, but shifted in phase, i.e., the emission from two poles of the pulsar. This conclusion could be confirmed by the detailed modeling of the pulse profiles similar to that performed by \citet{2003ApJ...590..424K}, which is, however, out of the scope of the current work.
Nevertheless, the complex phase dependence of spectral parameters again points to presence of an accretion column in \src, which implies then a super-critical luminosity $\sim10^{37}$\,erg\,s$^{-1}$ \citep{1976MNRAS.175..395B}, which only weakly depends on the assumed magnetic field strength of the neutron star unless it is magnetar-like \citep{2015MNRAS.447.1847M}. Comparing this luminosity with the observed flux yields a distance of $\sim10$\,kpc, in agreement with other considerations discussed above.
We note that a much more robust distance estimate can be obtained from follow-up observations of the counterpart.

\begin{figure}[t]
    \includegraphics[width=\columnwidth]{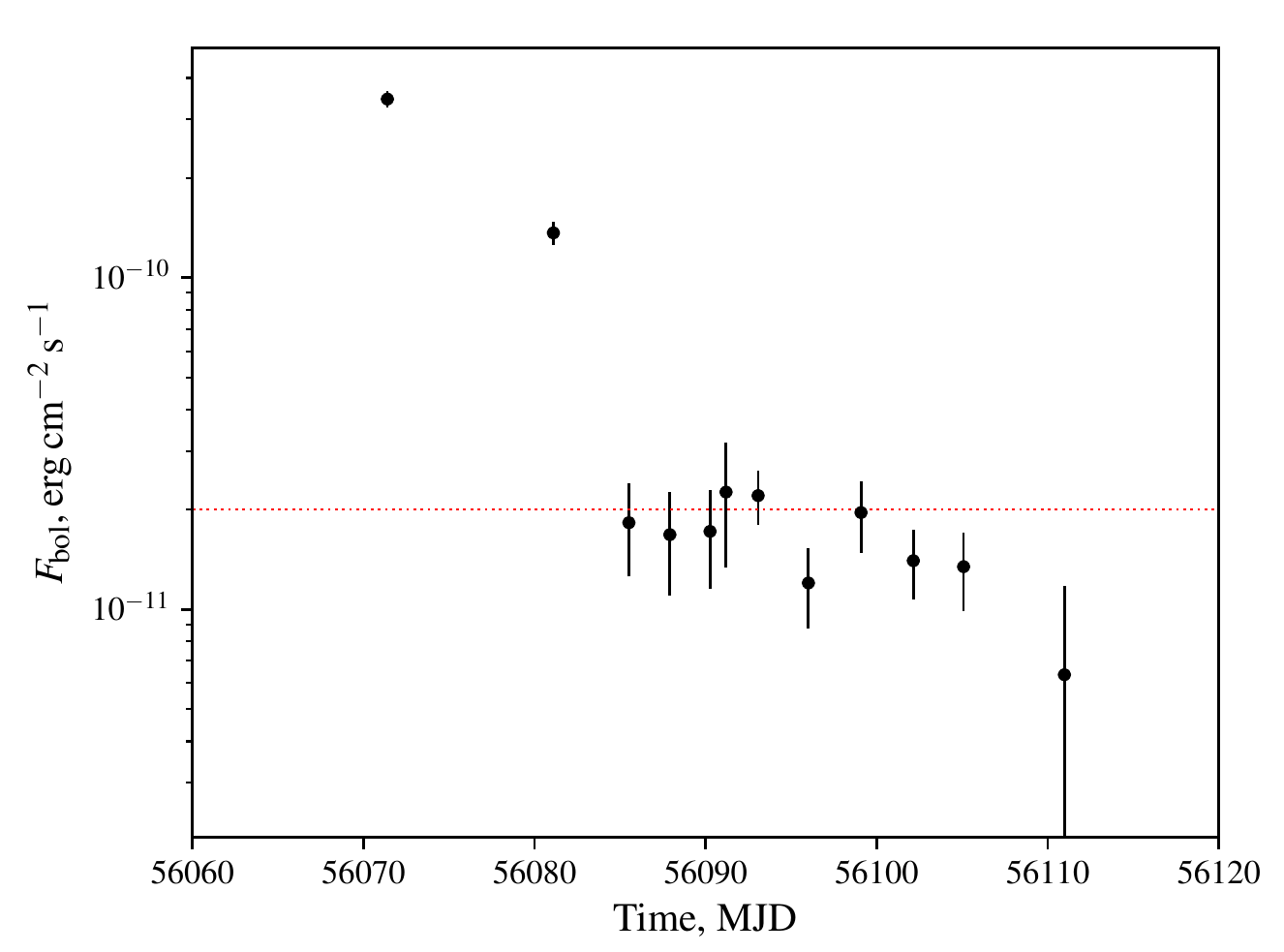}
        \caption{Bolometric source flux estimated based on historic \emph{Swift}/XRT lightcurve of the source as described in the text. The horizontal line indicates flux corresponding to transition to ``cold''-disc accretion regime.}
    \label{fig:histlc}
\end{figure}

The magnetic field of the source can not be directly estimated due to the lack of detection of any cyclotron line. Nevertheless, some indirect estimates can be done based on the analysis of archival \textit{Swift}/XRT data from the 2012 outburst. To estimate the bolometric source flux, we extracted spectra of each XRT observations and fitted them using the same model used for the \NS observations considering only the 0.5-10\,keV flux as a free parameter (i.e., multiplied the model by the \texttt{cflux} component). Here we used XRT data in the 0.5-10\,keV and 0.9-10\,keV bands for the photon counting and windowed modes respectively. Spectra were grouped to contain at least one count per energy bin. We used \texttt{lstat} to fit the spectra and estimate the flux uncertainties. The obtained fluxes were then multiplied by the bolometric correction of 4.96 estimated from the \NS fit to get an unabsorbed source flux in the 0.1-100\,keV energy band.
The resulting lightcurve of the declining tail of the 2012 outburst is shown in Fig.~\ref{fig:histlc}. 
The flux in the decay of the outburst levels out around MJD~56090 at $F_X\sim2\times10^{-11}$\,erg\,cm$^{-1}$\,s$^{-1}$ (note the logarithmic scale for flux), which is strikingly similar to the behaviour observed in several other pulsars
\citep{Tsygankov17_cold,2019A&A...621A.134T}. 

In particular, \cite{Tsygankov17_cold} argued that the flattening of the accretion rate at low fluxes observed in GRO~J1008$-$57 is attributed to the transition of the accretion disc to the non-ionized state with the lower viscosity, a well known mechanism in dwarf novae. The transition luminosity is defined by the temperature at the inner boundary of the accretion disc, which, for magnetized objects, is defined in turn by the magnetic field strength:
$$
L_{\rm cold}=9\times10^{33}k^{1.5}M_{1.4}^{0.28}R_6^{1.57}B_{12}^{0.86}{\rm erg\,s}^{-1}.
$$
Here $M_{1.4}$, $R_6$ and $B_{12}$ are neutron stars mass in units of 1.4$M_\odot$, 10$^{6}$\,cm and $10^{12}$\,G, respectively, and $k\simeq0.5$ is the coupling constant defining the effective magnetospheric radius with respect to the classical Alfv\'en radius.
As discussed by \cite{Tsygankov17_cold}, this estimate is rather crude, and depends on a number of assumptions, however, the scenario itself seems now to be confirmed by observations of the same state in several objects \citep{Tsygankov17_cold,2019A&A...621A.134T}.
We emphasize that as discussed by \cite{Tsygankov17_cold}, the transition to the cold-disc state is actually inevitable for slowly spinning pulsars with $P\gtrsim40$\,s. We therefore suggest that the observed flux leveling in \src corresponds to this transition, which can be used to estimate its magnetic field.

As already mentioned, the flux corresponding to the transition in \src is $F_X\sim2\times10^{-11}$\,erg\,cm$^{-1}$\,s$^{-1}$ implying a luminosity of $\sim2.4\times10^{35}(d/10{\rm\,kpc})^2$, which can be compared either with the theoretical value or observed transitional luminosity in the pulsar with the known field. In particular, in the source GRO~J1008$-$57 such a transition occurs at $\sim2\times10^{35}$\,erg\,s$^{-1}$ and the field strength is estimated at $8\times10^{12}$\,G \citep{2013ATel.4759....1Y} based on the observed cyclotron line energy. 
We emphasize that the strong absorption, arguments on the source luminosity and the pulse profile energy dependence strongly favor the large distance to the source (i.e., $\sim10$\,kpc as assumed above), which implies that the magnetic field in \src must also be
comparable to that in GRO~J1008$-$57, i.e., around $\sim10^{13}$\,G. This translates into a rather high value of the cyclotron line energy at $\gtrsim80$\,keV, beyond the observational range of \NS. We note that \textit{Insight}-HXMT covering a broader energy range 1-250\,keV \citep{2007NuPhS.166..131L,2019arXiv191009613Z} have also observed the source seven times between MJD~58773 to 58776 for 150~ks in total at the flux of $1.3-1.7\times10^{-9}$\,erg\,cm$^{-2}$\,s$^{-1}$, slightly higher than that observed by \NS. However, even then the source turned out to be too faint for the spectral analysis at hard X-rays. The timing analysis is ongoing and will be published elsewhere. We emphasize, however, that the source shall be the first priority for HXMT whenever it undergoes a brighter outburst as it is currently the only mission capable of testing presence of a cyclotron lines in the $\ge80$\,keV energy band.

\section*{Acknowledgements}
This research has made use of data and/or software provided by the High Energy Astrophysics Science Archive Research Center (HEASARC), which is a service of the Astrophysics Science Division at NASA/GSFC and the High Energy Astrophysics Division of the Smithsonian Astrophysical Observatory. This work made use of data supplied by the UK Swift Science Data Centre at the University of Leicester.
Authors thank the Russian Science Foundation (grant 19-12-00423), National Natural Science Foundation of China (NSFC U1838201),  German Academic Exchange Service (DAAD, project 57405000, VD), and the Academy of Finland travel grants 324550 (ST) and 316932 (AL) for support.
\bibliography{biblio}

\begin{thebibliography}{33}
\expandafter\ifx\csname natexlab\endcsname\relax\def\natexlab#1{#1}\fi

\bibitem[{{Arnaud}(1996)}]{1996ASPC..101...17A}
{Arnaud}, K.~A. 1996, in Astronomical Society of the Pacific Conference Series,
  Vol. 101, Astronomical Data Analysis Software and Systems V, ed. G.~H.
  {Jacoby} \& J.~{Barnes}, 17

\bibitem[{{Basko} \& {Sunyaev}(1976)}]{1976MNRAS.175..395B}
{Basko}, M.~M. \& {Sunyaev}, R.~A. 1976, \mnras, 175, 395

\bibitem[{{Deeter} {et~al.}(1981){Deeter}, {Boynton}, \&
  {Pravdo}}]{1981ApJ...247.1003D}
{Deeter}, J.~E., {Boynton}, P.~E., \& {Pravdo}, S.~H. 1981, \apj, 247, 1003

\bibitem[{Doroshenko \& Tsygankov(2019)}]{atel13208}
Doroshenko, V. \& Tsygankov, S. 2019, The Astronomer's Telegram

\bibitem[{{Doroshenko} {et~al.}(2018){Doroshenko}, {Tsygankov}, \&
  {Santangelo}}]{2018A&A...613A..19D}
{Doroshenko}, V., {Tsygankov}, S., \& {Santangelo}, A. 2018, \aap, 613, A19

\bibitem[{{Evans} {et~al.}(2009){Evans}, {Beardmore}, {Page}, {Osborne},
  {O'Brien}, {Willingale}, {Starling}, {Burrows}, {Godet}, {Vetere}, {Racusin},
  {Goad}, {Wiersema}, {Angelini}, {Capalbi}, {Chincarini}, {Gehrels}, {Kennea},
  {Margutti}, {Morris}, {Mountford}, {Pagani}, {Perri}, {Romano}, \&
  {Tanvir}}]{2009MNRAS.397.1177E}
{Evans}, P.~A., {Beardmore}, A.~P., {Page}, K.~L., {et~al.} 2009, \mnras, 397,
  1177

\bibitem[{{Harrison} {et~al.}(2013){Harrison}, {Craig}, {Christensen},
  {Hailey}, {Zhang}, {Boggs}, {Stern}, {Cook}, {Forster}, {Giommi},
  {Grefenstette}, {Kim}, {Kitaguchi}, {Koglin}, {Madsen}, {Mao}, {Miyasaka},
  {Mori}, {Perri}, {Pivovaroff}, {Puccetti}, {Rana}, {Westergaard}, {Willis},
  {Zoglauer}, {An}, {Bachetti}, {Barri{\`e}re}, {Bellm}, {Bhalerao},
  {Brejnholt}, {Fuerst}, {Liebe}, {Markwardt}, {Nynka}, {Vogel}, {Walton},
  {Wik}, {Alexander}, {Cominsky}, {Hornschemeier}, {Hornstrup}, {Kaspi},
  {Madejski}, {Matt}, {Molendi}, {Smith}, {Tomsick}, {Ajello}, {Ballantyne},
  {Balokovi{\'c}}, {Barret}, {Bauer}, {Blandford}, {Brandt}, {Brenneman},
  {Chiang}, {Chakrabarty}, {Chenevez}, {Comastri}, {Dufour}, {Elvis}, {Fabian},
  {Farrah}, {Fryer}, {Gotthelf}, {Grindlay}, {Helfand}, {Krivonos}, {Meier},
  {Miller}, {Natalucci}, {Ogle}, {Ofek}, {Ptak}, {Reynolds}, {Rigby},
  {Tagliaferri}, {Thorsett}, {Treister}, \& {Urry}}]{2013ApJ...770..103H}
{Harrison}, F.~A., {Craig}, W.~W., {Christensen}, F.~E., {et~al.} 2013, \apj,
  770, 103

\bibitem[{{HI4PI Collaboration} {et~al.}(2016){HI4PI Collaboration}, {Ben
  Bekhti}, {Fl{\"o}er}, {Keller}, {Kerp}, {Lenz}, {Winkel}, {Bailin},
  {Calabretta}, {Dedes}, {Ford}, {Gibson}, {Haud}, {Janowiecki}, {Kalberla},
  {Lockman}, {McClure-Griffiths}, {Murphy}, {Nakanishi}, {Pisano}, \&
  {Staveley-Smith}}]{2016A&A...594A.116H}
{HI4PI Collaboration}, {Ben Bekhti}, N., {Fl{\"o}er}, L., {et~al.} 2016, \aap,
  594, A116

\bibitem[{{Karino}(2007)}]{2007PASJ...59..961K}
{Karino}, S. 2007, \pasj, 59, 961

\bibitem[{Kennea(2019)}]{atel13219}
Kennea, J. 2019, The Astronomer's Telegram

\bibitem[{{Kennea} {et~al.}(2019{\natexlab{a}}){Kennea}, {Evans}, {Beardmore},
  {Bahramian}, {Krimm}, {Romano}, {Yamaoka}, {Serino}, \&
  {Negoro}}]{2019ATel13191....1K}
{Kennea}, J.~A., {Evans}, P.~A., {Beardmore}, A.~P., {et~al.}
  2019{\natexlab{a}}, The Astronomer's Telegram, 13191, 1

\bibitem[{{Kennea} {et~al.}(2019{\natexlab{b}}){Kennea}, {Evans}, {Beardmore},
  {Bahramian}, {Krimm}, {Romano}, {Yamaoka}, {Serino}, \&
  {Negoro}}]{2019ATel13195....1K}
{Kennea}, J.~A., {Evans}, P.~A., {Beardmore}, A.~P., {et~al.}
  2019{\natexlab{b}}, The Astronomer's Telegram, 13195, 1

\bibitem[{{Kraus} {et~al.}(2003){Kraus}, {Zahn}, {Weth}, \&
  {Ruder}}]{2003ApJ...590..424K}
{Kraus}, U., {Zahn}, C., {Weth}, C., \& {Ruder}, H. 2003, \apj, 590, 424

\bibitem[{{Krimm} {et~al.}(2012){Krimm}, {Kennea}, {Holland}, {Barthelmy},
  {Baumgartner}, {Cummings}, {Gehrels}, {Markwardt}, {Palmer}, {Sakamoto},
  {Skinner}, {Stamatikos}, {Tueller}, \& {Ukwatta}}]{2012ATel.4130....1K}
{Krimm}, H.~A., {Kennea}, J.~A., {Holland}, S.~T., {et~al.} 2012, The
  Astronomer's Telegram, 4130

\bibitem[{{Li}(2007)}]{2007NuPhS.166..131L}
{Li}, T.-P. 2007, Nuclear Physics B Proceedings Supplements, 166, 131

\bibitem[{{Lutovinov} \& {Tsygankov}(2009)}]{2009AstL...35..433L}
{Lutovinov}, A.~A. \& {Tsygankov}, S.~S. 2009, Astronomy Letters, 35, 433

\bibitem[{{Lutovinov} {et~al.}(2019){Lutovinov}, {Tsygankov}, {Karasev},
  {Molkov}, \& {Doroshenko}}]{2019MNRAS.485..770L}
{Lutovinov}, A.~A., {Tsygankov}, S.~S., {Karasev}, D.~I., {Molkov}, S.~V., \&
  {Doroshenko}, V. 2019, \mnras, 485, 770

\bibitem[{McCollum \& Laine(2019{\natexlab{a}})}]{atel13211}
McCollum, B. \& Laine, S. 2019{\natexlab{a}}, The Astronomer's Telegram

\bibitem[{McCollum \& Laine(2019{\natexlab{b}})}]{atel13222}
McCollum, B. \& Laine, S. 2019{\natexlab{b}}, The Astronomer's Telegram

\bibitem[{{Mushtukov} {et~al.}(2015){Mushtukov}, {Suleimanov}, {Tsygankov}, \&
  {Poutanen}}]{2015MNRAS.447.1847M}
{Mushtukov}, A.~A., {Suleimanov}, V.~F., {Tsygankov}, S.~S., \& {Poutanen}, J.
  2015, \mnras, 447, 1847

\bibitem[{{Negoro} {et~al.}(2019){Negoro}, {Yoneyama}, {Serino}, {Nakajima},
  {Maruyama}, {Aoki}, {Kobayashi}, {Mihara}, {Tamagawa}, {Matsuoka},
  {Sakamoto}, {Sugita}, {Nishida}, {Yoshida}, {Tsuboi}, {Iwakiri}, {Sasaki},
  {Kawai}, {Sato}, {Shidatsu}, {Kawai}, {Oeda}, {Shiraishi}, {Nakahira},
  {Sugawara}, {Ueno}, {Tomida}, {Ishikawa}, {Isobe}, {Shimomukai}, {Tominaga},
  {Ueda}, {Tanimoto}, {Yamada}, {Ogawa}, {Setoguchi}, {Yoshitake}, {Tsunemi},
  {Asakura}, {Ide}, {Yamauchi}, {Iwahori}, {Kurihara}, {Kurogi}, {Miike},
  {Kawamuro}, {Yamaoka}, {Kawakubo}, \& {Sugizaki}}]{2019ATel13189....1N}
{Negoro}, H., {Yoneyama}, T., {Serino}, M., {et~al.} 2019, The Astronomer's
  Telegram, 13189

\bibitem[{{Poutanen} {et~al.}(2013){Poutanen}, {Mushtukov}, {Suleimanov},
  {Tsygankov}, {Nagirner}, {Doroshenko}, \& {Lutovinov}}]{2013ApJ...777..115P}
{Poutanen}, J., {Mushtukov}, A.~A., {Suleimanov}, V.~F., {et~al.} 2013, \apj,
  777, 115

\bibitem[{{Saxton} {et~al.}(2008){Saxton}, {Read}, {Esquej}, {Freyberg},
  {Altieri}, \& {Bermejo}}]{2008A&A...480..611S}
{Saxton}, R.~D., {Read}, A.~M., {Esquej}, P., {et~al.} 2008, \aap, 480, 611

\bibitem[{Steele(2019)}]{atel13218}
Steele, I. 2019, The Astronomer's Telegram

\bibitem[{{Tanaka}(1986)}]{1986LNP...255..198T}
{Tanaka}, Y. 1986, in Lecture Notes in Physics, Berlin Springer Verlag, Vol.
  255, IAU Colloq. 89: Radiation Hydrodynamics in Stars and Compact Objects,
  ed. D.~{Mihalas} \& K.-H.~A. {Winkler}, 198

\bibitem[{{Tsygankov} {et~al.}(2017{\natexlab{a}}){Tsygankov}, {Doroshenko},
  {Lutovinov}, {Mushtukov}, \& {Poutanen}}]{2017A&A...605A..39T}
{Tsygankov}, S.~S., {Doroshenko}, V., {Lutovinov}, A.~A., {Mushtukov}, A.~A.,
  \& {Poutanen}, J. 2017{\natexlab{a}}, \aap, 605, A39

\bibitem[{{Tsygankov} {et~al.}(2019){Tsygankov}, {Doroshenko}, {Mushtukov},
  {Lutovinov}, \& {Poutanen}}]{2019A&A...621A.134T}
{Tsygankov}, S.~S., {Doroshenko}, V., {Mushtukov}, A.~A., {Lutovinov}, A.~A.,
  \& {Poutanen}, J. 2019, \aap, 621, A134

\bibitem[{{Tsygankov} {et~al.}(2016){Tsygankov}, {Lutovinov}, {Doroshenko},
  {Mushtukov}, {Suleimanov}, \& {Poutanen}}]{2016A&A...593A..16T}
{Tsygankov}, S.~S., {Lutovinov}, A.~A., {Doroshenko}, V., {et~al.} 2016, \aap,
  593, A16

\bibitem[{{Tsygankov} {et~al.}(2017{\natexlab{b}}){Tsygankov}, {Mushtukov},
  {Suleimanov}, {Doroshenko}, {Abolmasov}, {Lutovinov}, \&
  {Poutanen}}]{Tsygankov17_cold}
{Tsygankov}, S.~S., {Mushtukov}, A.~A., {Suleimanov}, V.~F., {et~al.}
  2017{\natexlab{b}}, \aap, 608, A17

\bibitem[{{Vybornov} {et~al.}(2018){Vybornov}, {Doroshenko}, {Staubert}, \&
  {Santangelo}}]{2018A&A...610A..88V}
{Vybornov}, V., {Doroshenko}, V., {Staubert}, R., \& {Santangelo}, A. 2018,
  \aap, 610, A88

\bibitem[{{Wilms} {et~al.}(2000){Wilms}, {Allen}, \&
  {McCray}}]{2000ApJ...542..914W}
{Wilms}, J., {Allen}, A., \& {McCray}, R. 2000, \apj, 542, 914

\bibitem[{{Yamamoto} {et~al.}(2013){Yamamoto}, {Mihara}, {Sugizaki}, {Sasano},
  {Makishima}, \& {Nakajima}}]{2013ATel.4759....1Y}
{Yamamoto}, T., {Mihara}, T., {Sugizaki}, M., {et~al.} 2013, The Astronomer's
  Telegram, 4759, 1

\bibitem[{{Zhang} {et~al.}(2019){Zhang}, {Li}, {Lu}, {Song}, {Xu}, {Liu},
  {Chen}, {Cao}, {Bu}, {Cai}, {Chang}, {Chen}, {Chen}, {Chen}, {Chen}, {Chen},
  {Chen}, {Cui}, {Cui}, {Deng}, {Dong}, {Du}, {Fu}, {Gao}, {Gao}, {Gao}, {Ge},
  {Gu}, {Guan}, {Gungor}, {Guo}, {Han}, {Hu}, {Huang}, {Huang}, {Huo}, {Jia},
  {Jiang}, {Jiang}, {Jin}, {Jin}, {Kong}, {Li}, {Li}, {Li}, {Li}, {Li}, {Li},
  {Li}, {Li}, {Li}, {Li}, {Li}, {Liang}, {Liao}, {Liu}, {Liu}, {Liu}, {Liu},
  {Liu}, {Liu}, {Liu}, {Lu}, {Lu}, {Luo}, {Luo}, {Ma}, {Meng}, {Nang}, {Nie},
  {Ou}, {Qu}, {Sai}, {Shang}, {Shen}, {Song}, {Sun}, {Tan}, {Tao}, {Tao},
  {Tuo}, {Wang}, {Wang}, {Wang}, {Wang}, {Wang}, {Wang}, {Wen}, {Wu}, {Wu},
  {Xiao}, {Xiao}, {Xiong}, {Xu}, {Yan}, {Yang}, {Yang}, {Yang}, {Yi}, {Yu},
  {Yuan}, {Zhang}, {Zhang}, {Zhang}, {Zhang}, {Zhang}, {Zhang}, {Zhang},
  {Zhang}, {Zhang}, {Zhang}, {Zhang}, {Zhang}, {Zhang}, {Zhang}, {Zhang},
  {Zhang}, {Zhang}, {Zhang}, {Zhang}, {Zhang}, {Zhang}, {Zhao}, {Zhao}, {Zhao},
  {Zheng}, {Zhu}, {Zhu}, {Zhuang}, \& {Zou}}]{2019arXiv191009613Z}
{Zhang}, S., {Li}, T., {Lu}, F., {et~al.} 2019, arXiv e-prints,
  arXiv:1910.09613

\end{thebibliography}
\end{document}